\documentclass[apl,twocolumn,superscriptaddress,showpacs]{revtex4}
\usepackage{amsmath,amssymb,graphicx,color,bm,epstopdf}

\begin{document}
\title{Rashba plasmon-polaritons in semiconductor heterostructures}

\author{I. V. Iorsh}
 \affiliation{National Research University of Information Technologies, Mechanics and Optics (ITMO), St.~Petersburg 197101, Russia}

\author{V. M. Kovalev}
\affiliation{Institute of Semiconductor Physics, Siberian Branch, Russian Academy of Sciences,
pr. Akademika Lavrent’eva 13, Novosibirsk, 630090 Russia}
\affiliation{Novosibirsk State Technical University, Novosibirsk, 630095 Russia}

\author{M. A. Kaliteevski}
\affiliation{Academic University - Nanotechnology Research and
Education Centre, Khlopina 8/3, 194021, St.Petersburg, Russia}
\affiliation{Ioffe Physical-Technical Institute, Polytekhnicheskaya
26, 194021, St.Petersburg, Russia}

\author{I. G. Savenko}
 \affiliation{Academic University - Nanotechnology Research and
Education Centre, Khlopina 8/3, 194021, St.Petersburg, Russia}
\affiliation{Science Institute, University of Iceland, Dunhagi-3, IS-107, Reykjavik, Iceland}
\affiliation{Division of Physics and Applied Physics, Nanyang Technological University 637371, Singapore}

\date{\today}

\begin{abstract}
We propose a concept of surface plasmon-polariton amplification in the structure comprising interface between dielectric, metal and asymmetric quantum well. Due to the Rashba spin-orbit interaction, mimina of dispersion relation for electrons in conduction band are shifted with respect to the maximum of dispersion dependence for holes in $\Gamma$-point. When energy and momentum intervals between extrema in dispersion relations of electrons and holes match dispersion relation of plasmons, indirect radiative transition can amplify the plasmons; excitation of leaky modes is forbidden due to the selection rules. Efficiency of the indirect radiative transition is calculated and design of the structure is analysed.
\end{abstract}


\maketitle


Surface plasmon-polaritons (SPPs), excitations which can appear and propagate along the boundary of a dielectric (insulator) and plasma, are well-studied quasiparticles, which were discovered nearly half a century ago \cite{Stern,Ritchie}. In recent years, surface plasmons have attracted substantial interest due to their potential applications in various optoelectronic devices \cite{Barnes,Coyle,Maier,Fedyanin2012,Briscoe2013,Zhang2013}.
The oscillation of an electron gas relative to much more massive ions with ``plasma angular frequency'' $\omega_p=\sqrt{\frac{4\pi Ne^2}{m}}$
(where $e$, $m$, and $N$ are the charge, mass and concentration of electrons) results in a frequency dependence of the dielectric constant of plasma: $\epsilon(\omega)=1-\frac{\omega_p^2}{\omega(\omega+i\gamma)}$, describing propagation of transverse electromagnetic waves (where $\gamma$ is the damping rate accounting for the dissipation of plasma oscillation energy and $\omega$ is the angular frequency \cite{Stix,Aliev,Montgomery,Jackson}).

For the most of metals, plasma frequency corresponds to the ultraviolet light \cite{Stix} and $\gamma\ll\omega_p$. For instance, in gold $\hbar\omega_p=7.9$ eV and $\hbar\gamma=0.09$ eV at room temperature.
In metallic metamaterials spectral dependence of dielectric constant is characterized by the effective plasma frequency, which can be much lower than the one that belongs to the bulk metal \cite{Pendry,Brand}.
Plasma frequency of free electrons in semiconductors is determined by the doping level and the effective mass of the relevant band. In solids generally, the damping rate can be substantially reduced by lowering the temperature. Thus, for all cases, a collisionless plasma for which $\gamma=0$ is quite an approximation to a system which is attainable for many applications.
\begin{figure}[!htb]
\includegraphics[width=1.0\linewidth]{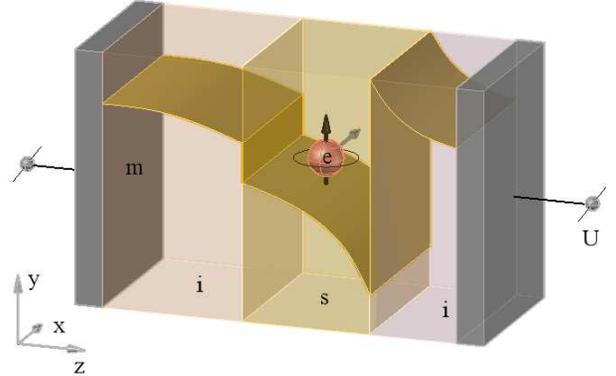}
\caption{Sketch of the system: metal (m) -insulator (i) -semiconductor (s) structure with asymmetric QW in the middle (a layer denoted as 's'). Some of the semiconductor layers (buffer layer, cladding layers and emitters) are skipped in figure for the clarity reasons. Due to the Rashba spin-orbit interaction, resulting from the asymmetry of the QW, the system can serve as a source of surface plasmons on the metal-dielectric boundary, controlled by the bias voltage $U$.}
\label{fig:1}
\end{figure}

However, in realistic device, SPPs have finite propagation length due to absorbtion in metal, and their pumping is required.
For the existing schemes of puping, light can be emitted not only into a surface plasmon mode but also into leaky light modes within the light cone in various directions \cite{Walters2010}. It results in reduction of efficiency of pumping. Moreover, in the case when the radiative recombination occurs for the electron lying not in the band minima, more efficient non-radiative processes such as acoustic phonon scattering would preclude the efficient radiative recombination.
Current Letter is aimed at the development of design of a device, in which energy is pumped into the SPP mode only.
This is achieved by means of an effective shift of the conductance band minimum towards higher $k$-states in reciprocal space. Then, the semiconductor becomes indirect one and optical transitions turn to be damped due to the energy and momentum conservation laws. On the other hand, the transitions accompanied by the SPPs emission still remain possible from this viewpoint. In our work, the Rashba spin-orbit interaction (RSOI) effect is used to shift the minimum of the conduction band, that allows to suppress optical leaky modes but does not hinder the SPP emission.


Consider an asymmetric quantum well (QW) buried in a semiconductor structure covered with a metallic layer, as shown in Fig.~1. If the QW is optically or electrically pumped, the top states of the valence band are populated with holes, while the bottom of the conduction band is populated with electrons, and an inversion of population occurs.

\begin{figure}[!htb]
\includegraphics[width=0.7\linewidth]{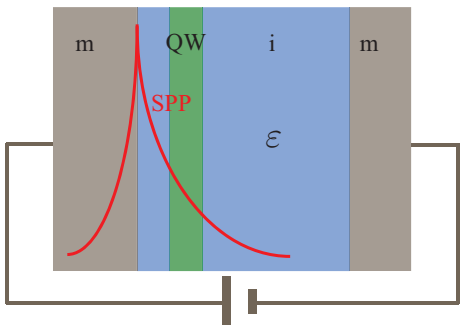}
\caption{Schematic illustration of the SPP mode penetration into neighbouring layers (metal $m$, semiconductor QW and the insulators $i$) and overlap of the SPP electric field profile with the QW region. The depicted structure resembles one presented in Fig.~1 (in cross-section).}
\label{fig:2}
\end{figure}

It is well known, that the dispersion relation of conduction electrons in an asymmetric QW has linear in
$\textbf{p}$ terms coupled to the electron spin \cite{Winkler}. This mechanism of spin-orbit coupling (widely called Rashba spin-orbit interaction) can be characterized by the interaction Hamiltonian \cite{Rashba1984}:
\begin{eqnarray}
{\cal H}_{SO}(\textbf{p})=\alpha [\textbf{p}\times\sigma]\textbf{n},
\end{eqnarray}
where $\alpha$ is a strength of the RSOI, $\textbf{p}$ is
the electron momentum in the plane of the QW, $\sigma$ are the Pauli
matrices and $\textbf{n}$ is a normal to the QW plane vector.
Part of full system Hamiltonian related to the QW electrons reads:
\begin{eqnarray}{{\cal H}}_e=\frac{\textbf{p}^2}{2m_e}+{\cal H}_{SO}(\textbf{p}).
\label{EqElHam}
\end{eqnarray}
The solution of the eigenvalue problem for the Hamiltonian \eqref{EqElHam} takes the form:
\begin{eqnarray}
\label{EqRashEigen}
E_\mu(\textbf{p})&=&\frac{\mathbf{p}^2}{2m^\ast}+\mu\alpha p,\\
\nonumber \Psi_{\mu}(\mathbf{p})&=&\frac{1}{\sqrt{2}}\left(%
\begin{array}{c}
  -i\mu\frac{p_+}{p} \\
  1 \\
\end{array}%
\right)e^{i\textbf{pr}},\\
\nonumber
&&p_+=p_x+ip_y,
\end{eqnarray}
where $p=|\mathbf{p}|$; $\mu=\pm 1$ is the electron chirality~\cite{Rashba1984} (number of the energy
dispersion branch). From \eqref{EqRashEigen} it becomes clear that in presence of RSOI, dispersion relation for electrons in conduction band are spin-dependent and has minima shifted in respect to $\Gamma$-point and positioned in the points $p=\pm\alpha m_e/\hbar$  (see Fig.~3).

Full Hamiltonian of the system can be written in form:
\begin{equation}
{\cal H}={\cal H}_e+{\cal H}_P,
\end{equation}
where the second term
\begin{eqnarray}
{\cal H}_P=V_p\sum_{k,k_p}({a_k^{C\dagger}}a_{k-k_p}^Va_{k_p}^P + a_k^C{a_{k-k_p}^{V\dagger}}{a_{k_p}^{P\dagger}})
\end{eqnarray}
describes the possible indirect transition between the conduction and valence band states, accompanied by the surface plasmon emission. Here $a_k^{C(V,P) \dagger}$, $a_k^{C(V,P)}$ are creation-annihilation operators of the conduction band electrons, valence band holes and SPP, respectively.

The SPP can propagate along the interface and its electric field overlaps with the QW (see Fig.~2). Thus, the SPP mode can be amplified by the indirect radiative transitions. Note, that efficiency of the transition in this case is increased due to the Purcell effect, since the amplitude of electric field of SPP is higher near the interface.
\begin{figure}[!htb]
\includegraphics[width=1.0\linewidth]{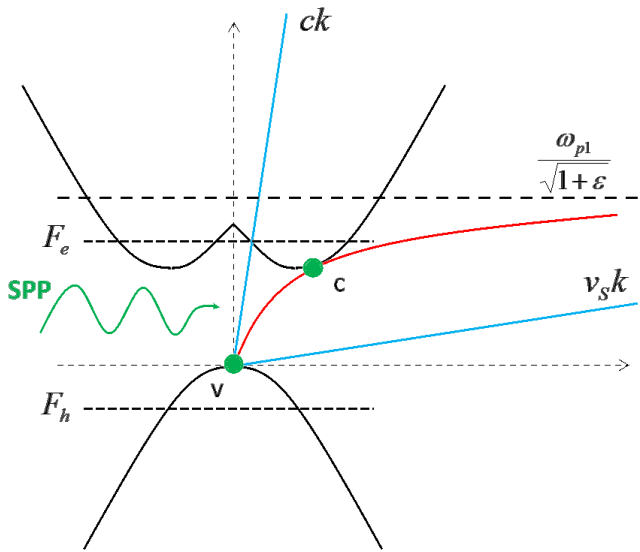}
\caption{Scheme of transitions in the system. The conduction band
minimum (point `c' in figure) is shifted in a certain wavevector
$|k_P|=m\alpha/\hbar^2$ due to the spatial inhomogenity of the QW.
The shift occurs due to the Rashba effect. It is shown here only
the lower branch of the electron dispersion \eqref{EqRashEigen}
with quantum number $\mu=-1$. The valence band shift is neglected,
thus the minimum of the valence band is located in the
$\Gamma$-point (`v'). The transition between the upper (conduction
band) state `c' and the lower (valence band) state `v'
($\Gamma$-minimum) can result in a SPP emission since the plasmon
dispersion (red line) matches both the points `c' and `v' and both
energy and momentum conservation laws are fully satisfied in such
a process. Blue curves correspond to the bare photon and acoustic
phonon dispersions. Dashed horizontal line illustrates the
high-$k$ asymptotics of the plasmon dispersion. Dashed lines with
keys $F_e,F_h$ represent quasi-Fermi levels of electrons and holes
respectively.} \label{fig:1}
\end{figure}
The corresponding matrix element $V_P$ can be estimated using a standard formula (we consider the matrix element $\textbf{k}$-independent for simplicity):
\begin{align}
V_{P}=\frac{2 q}{m_0c}\langle u_c|\mathbf{e}\mathbf{p}|u_v\rangle\int\limits_{d}^{d+d_{QW}}F^*_cA(z)F_vdz,
\end{align}
where $A(z)$ is the amplitude of the vector potential corresponding to the surface plasmon mode; $F_c$, $F_v$ are the envelope functions of the electron and hole in the QW, $d_{QW}$ is the QW thickness; $u_c$, $u_v$ are electron and hole Bloch functions respectively and $\mathbf{e}$ is the polarization vector associated with the plasmon mode. Integral over the Bloch functions can be evaluated as \cite{Galickiy}:
\begin{equation}
\frac{\langle u_c|\mathbf{e}\mathbf{p}|u_v\rangle}{m_0}\approx \sqrt{\frac{3E_g}{4\mu}},
\end{equation}
where the reduced mass $\mu$ is $\mu^{-1}=m_{e}^{-1}+m_{h}^{-1}$.
Before evaluating the integral over the envelope functions, we should choose the nomalization of the wavefunctions properly. Let us for simplicity assume infinite barrier quantum well solutions for electrons and holes. Normalization constant in this case equals $\sqrt{2/(d_{QW}S)}$, where $S$ is the area of the device. We normalize the magnetic potential $A$ applying the condition that total electromagnetic energy associated with the plasmonic mode should be equal to $\hbar\omega_P/(2\sqrt{1+\varepsilon})$. Using this condition, we can evaluate the value of magnetic potential amplitude at the metal-dielectric boundary  $|A_0|$:
\begin{align}
\frac{|qA_0|}{c}=\frac{q}{2}\sqrt{\frac{|k_z|\hbar\sqrt{1+\varepsilon}}{S\omega_P}},
\end{align}
where $k_z$ is the component of the wavevector perpendicular to the layers: $k_z=\sqrt{\varepsilon k_0^2-k_x^2}$; the absolute value of the plasmon wavevector in vacuum is $\left(E_e(\alpha m_e/\hbar^2)-E_h(0)\right)/c$, with $E_e$ and $E_h$ being the corresponding electron and hole energies. Due to the momentum conservation law, the in-plane component of the plasmon wavevector $k_x$ is equal to $\alpha m^*/\hbar^2$.
Thereby, the integral over the normalized envelope functions in the approximation of infinite barriers reads:
\begin{align}
\int\limits_{d}^{d+d_{QW}}F^*_cA(z)F_vdz \approx \frac{2\pi^2e^{-|k_z|d_{QW}}}{|k_z|d_{QW}(4\pi^2+|k_z|^2d_{QW}^2)}.
\end{align}
Finally, the matrix element can be evaluated as:
\begin{align}
V_P\approx \sqrt{\frac{|k_z|\hbar}{S\omega_P}}\sqrt{\frac{3E_g}{\mu}}\frac{q\pi^2e^{-|k_z|d_{QW}}(1+\varepsilon)^{1/4}}{|k_z|d_{QW}(4\pi^2+|k_z|^2d_{QW}^2)}.
\end{align}
This estimation allows to find the order of magnitude of the matrix element.


Let us consider a planar heterostructure based on the family of alloys $InAlGaAs$ with an asymmetric QW consisting of the particular alloy $(Ga_xIn_{1-x}As)_{z}(Al_yIn_{1-y}As)_{1-z}$ with $x=0.47$, $y=0.48$ and $z$ being a tuning parameter. The bandgaps ($E_g$), effective masses ($m_e$, $m_h$) and semiconductor dielectric constants ($\varepsilon_s$) of the constituents can be taken from \cite{Meyer,Ioffe}.
The surface area of the in-plane of the device equals to $S=10\times 10$ $\mu m^2$, the QW width $d_{QW}=10$ nm.

Due to the RSOI, the conduction band becomes split onto two subbands, and the minima of these subbands lie at $k_z\neq 0$, namely $k_z=\mp m_e\alpha/\hbar^2$, where the Rashba constant $\alpha$ was taken equal to some average characteristic value for the alloys which include $In$, $\alpha=1.2\cdot 10^{-9}$ $eV\cdot cm$ \cite{Silsbee,Silva}. Thus, we obtain $k_z\approx 10^7$ $m^{-1}$.
We choose $Pt$ for the material of the metal layer. Then the plasmon frequency reads $\omega_p\approx 5.1$ $eV$ \cite{DrudeWeb1,DrudeWeb2}, and one immediately calculates the dielectric constant of the metal using the Drude model, $\varepsilon_m=1-\omega_p^2/\omega^2$, where $\omega=\tilde{E}_g=E_g-\hbar^2k_z^2/2m_e$.

Further, one should solve the equation
\begin{eqnarray}
E_g(x,y,z,T,k_z)&=&\hbar\omega(k_z)\\
\nonumber
&=&c\hbar k_z\sqrt{\frac{\varepsilon_s+\varepsilon_m}{\varepsilon_s\varepsilon_m}}\approx\frac{\hbar\omega_p}{\sqrt{1+\varepsilon_s}},
\end{eqnarray}
where the dielectric constants are frequency-dependent, $\varepsilon_{s(m)}=\varepsilon_{s(m)}(\omega)$, and find the parameters of the structure at which the energy conservation law is satisfied in the process of plasmon emission. Thus, at $2$ $K$ the composition reads $z=0.86$.

Further, to estimate the possible operation temperature, we compare the energy $\hbar^2k_z^2/2m^*$ with $k_BT$ and conclude that in the chosen materials the ratio is close to unity. It should be noted, that, in principle, the temperature can be higher than $1-2$ $K$ if the RSOI constant is large enough and $k_z$ approaches $10^8$ m$^{-1}$. This latter value is the cut-off momentum for most of metals \cite{Rocca,Politano,Pitarke,Johnson}, since as long as a plasmon wavevector becomes higher, the losses grow exponentially and the mean path of the particle becomes negligible.
It is important to note, that the RSOI strength can be controlled by the external voltage \cite{Nitta,Lommer}. By means of this factor, it becomes possible to control the plasmon emission, switching it on and off.

We have considered a system of coupled conductance electrons, valence band holes  and surface plasmon polaritons in semiconductor microcavity and showed that the Rashba spin-orbit interaction results in possibility to make a source and amplifier of surface plasmons in semiconductor-based heterostructures.

The authors thank I. A. Shelykh, M. M. Glazov and O. V. Kibis for useful discussions.
This work was supported by Rannis ``Center of excellence in polaritonics'', FP7 IRSES projects ``SPINMET'', ``POLAPHEN'' and ``POLALAS'', Russian
Fund of Basic Research and COST ``POLATOM'' program. I.G.S. acknowledges support of the Eimskip foundation.


\end{document}